\documentclass[fleqn,usenatbib,useAMS]{mnras}
\usepackage[utf8]{inputenc} 

\usepackage[T1]{fontenc}
\usepackage{ae,aecompl}

\usepackage{graphicx} 
\usepackage{amsmath}  
\usepackage{txfonts}  
\usepackage{bm}       

\title[Starspot induced effects in microlensing]{Starspot induced effects in
  microlensing events with rotating source star}

\author[M. Giordano et al.]
{M. Giordano$^{1,2}$\thanks{E-mail: mose.giordano@le.infn.it},
  A. A. Nucita$^{1,2}$, F. De Paolis$^{1,2}$ and G. Ingrosso$^{1,2}$\\
  $^{1}$Dipartimento di Matematica e Fisica `\emph{E. De Giorgi}',
  Universit\`{a} del Salento, Via per Arnesano, CP 193, I-73100 Lecce, Italy \\
  $^{2}$INFN, Sezione di Lecce, Via per Arnesano, CP 193, I-73100 Lecce, Italy}

\date{Accepted XXX. Received YYY; in original form ZZZ}

\pubyear{2015}

\begin{document}
\label{firstpage}
\pagerange{\pageref{firstpage}--\pageref{lastpage}}
\maketitle

\begin{abstract}
  We consider the effects induced by the presence of hot and cold spots on the
  source star in the light curves of simulated microlensing events due to either
  single or binary lenses taking into account the rotation of the source star
  and the orbital motion of the lens system.  Our goal is to study the anomalies
  induced by these effects on simulated microlensing light curves.
\end{abstract}




\section{Introduction}
\label{sec:introduction}

Gravitational microlensing has proven to be an exceptional tool to investigate
several astrophysical phenomena.  It is most often used to probe the lens system
(located between the source and the observer) that may be a star, possibly with
its planetary system, a MACHO (Massive Astrophysical Compact Halo Object) or
even a free-floating planet.  Moreover, gravitational microlensing behaves as a
powerful natural telescope that allows to investigate the source star located at
kpc distance from Earth (see, e.g.,~\citealt{2001PASP..113..903G}).  The source
star can be best studied when it crosses the caustics and moreover large finite
source effects are present, that is when the projection of the source disc on to
the lens plane is sufficiently large.  This happens when the lens and/or the
source are close enough or the source is a red giant star.

The main applications of gravitational microlensing to stellar physics are the
study of the limb-darkening of source stars~\citep{1995ApJ...449...42W} and of
irregularities, like cold and hot spots, on their surface.
\citet{2000ApJ...529...69H} and \citet{2002MNRAS.335..539H} studied the lensing
of a spotted star by a single lens, \citet{2000MNRAS.316..665H} and
\citet{2002MNRAS.335..195C} later extended the study to the binary lens case.
However, none of these papers has taken into account the possibility that the
source star and/or the binary lens system rotate.  The aim of the present work
is that of studying the anomalies induced by these effects on simulated
microlensing light curves.

\section{Event simulation}
\label{sec:event-simulation}

In exoplanet search it is well known that the presence of stellar spots can
induce features either in the observed lightcurves and in the radial velocity
profiles, and may mimic the signal due to one or more planets orbiting around
the star~\citep{2001A&A...379..279Q}.  Indeed, recently that effect generated a
false identification of a planet in the initial analysis of the microlensing
event MOA-2010-BLG-523~\citep{2013ApJ...763..141G}, as due to the intrinsic
variability of the source star.

This has pushed us to consider in detail the effects on simulated microlensing
events of one or more starspots on the surface of the source star.

As usual in describing microlensing events, we use use the Einstein radius
\(R_{\textup{E}}\) and the Einstein time \(t_{\textup{E}}\) as the typical
scales for lengths and times.  They are defined as follows:
\begin{align}
  R_{\textup{E}} &= \sqrt{\frac{4GM}{c^{2}}
    \frac{D_{\textup{ls}} D_{\textup{l}}}{D_{\textup{s}}}} \\
  t_{\textup{E}} &= \frac{R_{\textup{E}}}{v_{\perp}}
\end{align}
where \(M\) is the total mass of the lens system, \(D_{\textup{l}}\) is the
distance from the observer to the lens, \(D_{\textup{s}}\) is the distance from
the observer to the source star, \(D_{\textup{ls}} = D_{\textup{s}} -
D_{\textup{l}}\) is the distance between the lens system and the source star,
\(v_{\perp}\) is the speed of the source relative to the lens, perpendicular to
the line of sight.  In the following, all lengths and times are expressed in
units of Einstein radius and Einstein time respectively.

We calculate the amplification \(A_{\textup{s}}\) of the spotted star as the
weighted average of the amplification \(A(\bm{r})\) over the star disc
\(\mathcal{S}\), using as weight the surface brightness \(f(\bm{r})\), which
will be defined below,
\begin{equation}
  \label{eq:ampl}
  A_{\textup{s}} = \frac{\int_{\mathcal{S}} A(\bm{r})
    f(\bm{r})\mathop{}\!\textup{d}^{2}\bm{r}}{\int_{\mathcal{S}}
    f(\bm{r})\mathop{}\!\textup{d}^{2}\bm{r}}.
\end{equation}
The amplification \(A(\bm{r})\) can be either the amplification for a single
lens (see, e.g.,~\citealt{1986ApJ...304....1P}),
\begin{equation}
  A(u) = \frac{u^{2} + 2}{u\sqrt{u^{2} + 4}}
\end{equation}
being \(u\) the projected distance between the lens and the source in units of
Einstein radius, or the amplification induced by a binary
lens~\citep{2012RAA....12..947M}.  In the latter case, we calculate the
amplification of the source by combining two different methods:
\begin{itemize}
\item inverse ray-shooting~\citep{1986A&A...164..237S,1986A&A...166...36K} when
  the source is close to the caustics;
\item \citet{1995ApJ...447L.105W} method when the source is far enough from the
  caustics.  In particular, in order to solve the fifth-order polynomial
  equation we employed the root finder by~\citet{2012arXiv1203.1034S}.
\end{itemize}
The surface brightness profile \(f(\bm{r})\) in equation~\eqref{eq:ampl} is
defined as
\begin{equation}
  f(\bm{r}) =
  \begin{cases}
    l(\bm{r})   & \text{outside the spot} \\
    c~l(\bm{r}) & \text{inside the spot}
  \end{cases}
\end{equation}
where \(l(\bm{r})\) is the brightness of the unspotted star, and \(c\) is the
\emph{contrast parameter}, that is the ratio between the brightness of the spot
and the unspotted surface.  The case \(c>1\) corresponds to a hot spot, \(c<1\)
is for cold spots.  To model the source star surface brightness, we adopt the
linear limb-darkening law~\citep{2000ApJ...532..340A}
\begin{equation}
  l(\bm{r}) = l(r) = \Biggl[1 - \Gamma_{\lambda}\Biggl(1 - \frac{3}{2}\sqrt{1 -
    \frac{r^{2}}{\rho^{2}}}\Biggr)\Biggl]\tilde{F},
\end{equation}
where \(r\) is the distance from the centre of the star disc,
\(\Gamma_{\lambda}\) is the limb-darkening coefficient at the wavelength
\(\lambda\), \(\rho\) is the radius of the source, and \(\tilde{F} =
F/\upi\rho^{2}\) is the total flux \(F\) of the star averaged over the stellar
surface.  The star spot modellization is described in
appendix~\ref{sec:modelling-star-spots}.  Integration over the star disc is
carried out by using the \textsc{Cuba} library, developed
by~\citet{2005CoPhC.168...78H}.  Since the stellar spots can be quite small, we
have to be sure that the region around each spot is well sampled.  Thus we chose
in particular the Divonne algorithm, since it can allow us to increase sampling
around the position of given points of the integrand function, the starspots in
our case.

Differently from previous works, in this paper we consider also the orbital
motion of the lens system and the rotation of the source star.  In the next
section, we present the results of the simulated microlensing events to
highlight the effects of source spinning and binary lens orbital motion.

\section{Simulations}
\label{sec:simulations}

\subsection{Single lens}
\label{sec:single-lens}

\begin{figure}
  \centering
  \includegraphics[width=\columnwidth]{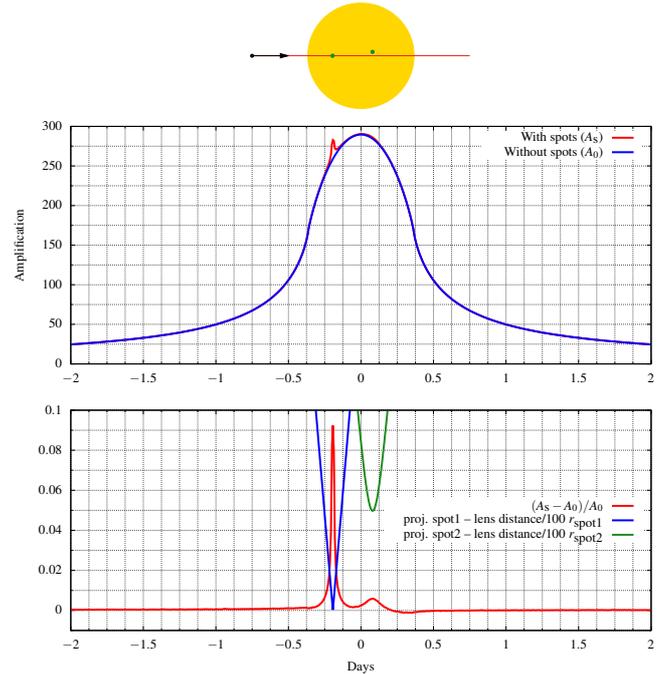}
  \caption{Simulations of a microlensing event by a single lens.  In the upper
    panel the source, with its hot spots, and the trajectory of the lens are
    represented.  In the middle panel it is shown the amplification of the
    source with and without the spots.  In the lower panel the fractional
    deviation of the amplification from that of a spotless event
    \((A_{\textup{s}} - A_{0})/A_{0}\) and the projected distance between the
    lens and each spot are shown.}
  \label{fig:single-lens}
\end{figure}
In Fig.~\ref{fig:single-lens} we show the results of a simulation for the single
lens case.  We set the ratio \(D_{\textup{l}}/D_{\textup{s}}\) between the
distance to the lens and to the source to be \(0.5\); the source projected
radius \(\rho=0.0075\) and its rotation period is \(0.61~t_{\textup{E}}\).  The
distance of closest approach between the point-like lens and the centre of the
source is \(u_{0} = 0\).  The two hot spots on the source have radius
\(R_{\textup{spots}} = 0.03~\rho\) and contrast parameter \(c = 4\).  One of the
spots has latitude \(\theta_{1} = 0\), and so the lens passes right over it; the
other spot has latitude \(\theta_{2} = \upi/21\).

It should be noted that, in this case, even if the star had a shorter period,
the probability to see other spot-induced peaks would be negligible.  In
particular, it would be zero if the lens does not move parallel to the equator
of the source, since the projected distance between the lens and the spot must
be as small as possible for the spot-induced peak to be visible (see the lower
panel of Fig.~\ref{fig:single-lens}).

\subsection{Binary lens}
\label{sec:binary-lens}

\begin{figure*}
  \centering
  \includegraphics[width=0.4\textwidth]{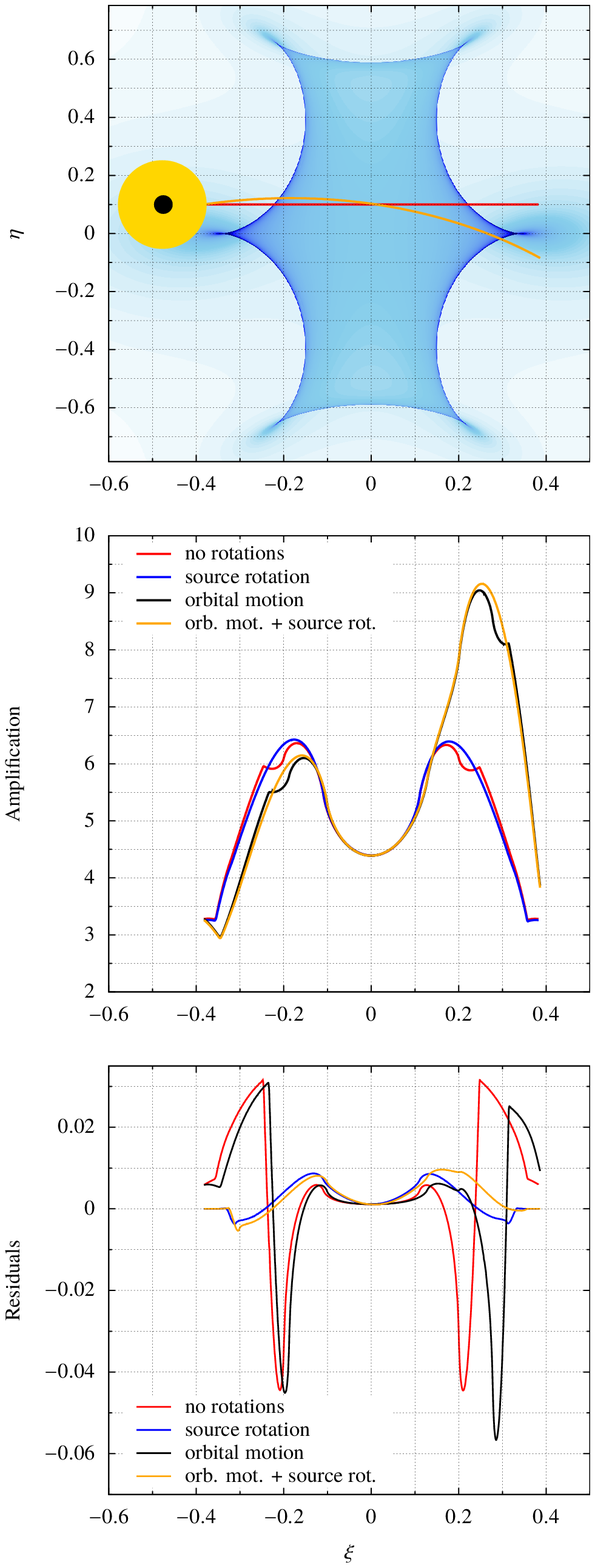}
  \qquad{}\qquad{}
  \includegraphics[width=0.4\textwidth]{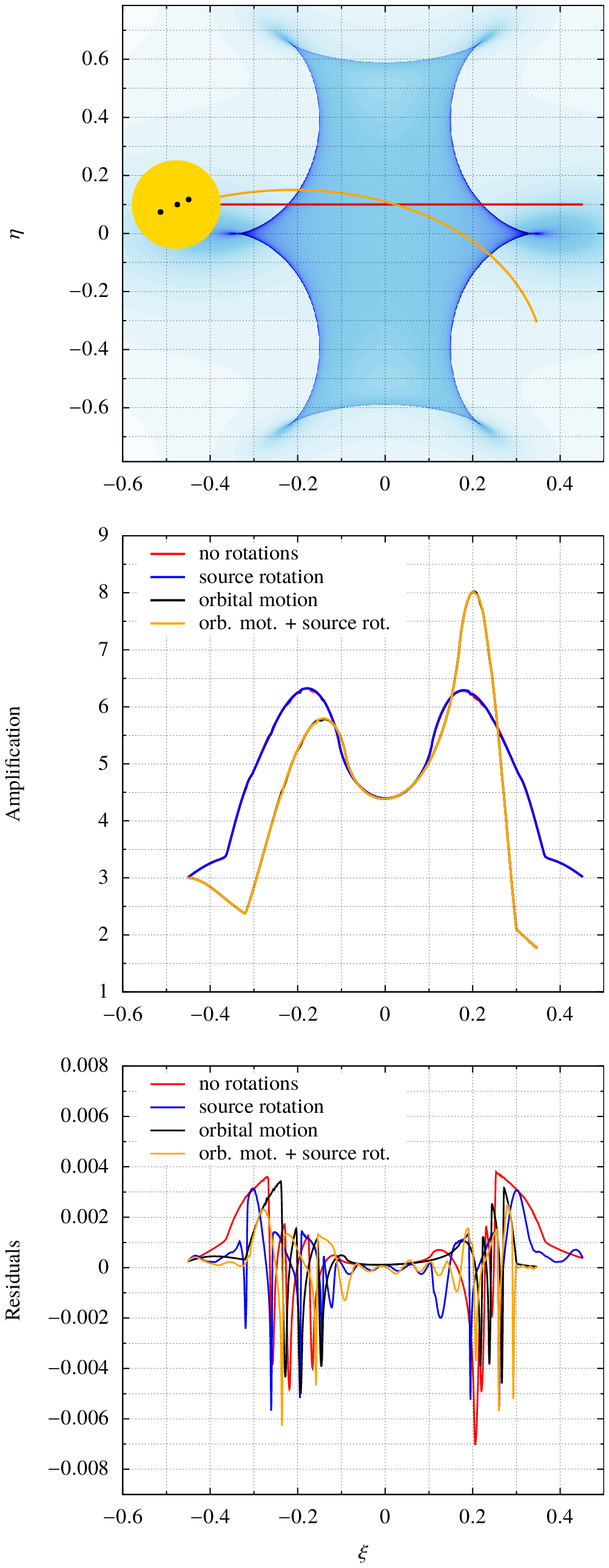}
  \caption{Simulations of microlensing events by a binary system.  We simulated
    events with and without orbital motion of the lenses and rotation of the
    source itself.  In the upper panels, the amplification map and the
    trajectories of the source in the lens plane \(\xi-\eta\) are shown in the
    case of no orbital motion of the system (red straight line) and if the lens
    system rotates (gold line).  In the middle panels there are the
    amplification curve for the different trajectories and in the lower panels
    the residuals between the case of a spotted source and the unspotted one are
    reported.  In the left-hand panels, a source star with one coldspot on its
    surface is considered, while in the right-hand panels, we show the results
    for a source star with five coldspots (only three initially on the visible
    side).  We refer to the text for the parameters used in these simulations.}
  \label{fig:binary-lens}
\end{figure*}
In Fig.~\ref{fig:binary-lens} our simulation results for two binary lens cases
are shown.  For simplicity, we assumed the orbit of the lens systems to have
eccentricity \(e = 0\) and to be face-on.  The distance of closest approach
between the source and the centre of mass of the lenses is \(u_{0} = 0.1\).  The
source trajectories are shown in the upper panels of Fig.~\ref{fig:binary-lens}.

For comparison purposes, in the left-hand panel of Fig.~\ref{fig:binary-lens} we
choose the system parameters of the event simulated to be quite close to the
binary events simulated by~\cite{2000MNRAS.316..665H}.  The lens system is
constituted by two equal mass objects, so their mass ratio is \(q = 1\), and a
projected separation of \(b = 1\).  The ratio \(D_{\textup{l}}/D_{\textup{s}}\)
is equal to \(0.5\).  The orbital period, when non null, is set to
\(10.1~t_{\textup{E}}\).  The source star has a projected radius of \(\rho =
0.1\) and its rotation period is \(1.23~t_{\textup{E}}\).  The limb-darkening
coefficient used in the simulations is \(\Gamma_{\lambda} = 0.45\).  The spot,
with centre on the equator, is a cold spot with \(c = 0.1\) and radius equal to
\(0.2~\rho\).

In the right-hand panel of Fig.~\ref{fig:binary-lens}, we consider a simulated
event with more realistic parameters, in particular smaller spots and a larger
contrast parameter \(c\), similarly to the real stellar spots detected with the
transit method (see, e.g.,~\citealt{2014MNRAS.443.2391M,2009A&A...504..561W}).
In this case the residuals are necessarily smaller, but the effects are
nevertheless interesting to be studied.  The lens system components have the
same mass (\(q = 1\)), and projected separation \(b = 1\).  Here, the ratio
\(D_{\textup{l}}/D_{\textup{s}}\) has been fixed to \(7/8\).  When the orbital
motion of the lenses is active, the orbital period is \(6~t_{\textup{E}}\).  The
source star has a rotation period of \(0.15~t_{\textup{E}}\), and a projected
radius of \(\rho = 0.1\).  The limb-darkening coefficient used in the simulation
is \(\Gamma_{\lambda} = 0.45\).  In this simulations, there are five cold spots
with \(c = 0.25\) and radius equal to \(0.05\rho\).  Three of the spots are on
the same side of the source sphere, with latitude \(\theta_{1} = 0\),
\(\theta_{2} = -\upi/12\), \(\theta_{3} = \upi/18\), and initial longitude
\(\varphi_{1} = 0\), \(\varphi_{2} = -\upi/8\), \(\varphi_{3} = \upi/12\)
respectively.  The other two spots are on the opposite side, so they cannot be
seen in Fig.~\ref{fig:binary-lens}, have latitude \(\theta_{4} = \theta_{5} =
0\), and initial longitude \(\varphi_{4} = 2\upi/3\), and \(\varphi_{5} =
4\upi/3\).

\section{Results and discussion}
\label{sec:discussion}

The crossing of a caustic by a spotted star in microlensing events introduces
extra peaks or dips in the amplification curve, which may be confused with
planetary signatures.  The spot-induced features are always very close to the
main peak, because the secondary peaks or dips must fall within the angular size
of the source.  As already noted by~\citet{2000MNRAS.316..665H}, the probability
to detect stellar spots is higher in the binary lens case than for the single
lens.  So, we simulated microlensing events by single and binary lens systems of
sources with hot and/or cold spots. We also took into account the spin of the
source star and the orbital motion of the lens binary.  Indeed, we emphasize
that the probability of detecting starspots tends to increase if the source
rotates substantially during the lensing events (especially in the binary lens
case) since the caustics are extended.

The repetition of the same feature in the blue and gold residuals of the
simulation in the right-hand panel of Fig.~\ref{fig:binary-lens} could, in
principle, make it possible to estimate the rotation period of the source with a
generalized Lomb-Scargle periodogram~\citep{2009A&A...496..577Z}.\footnote{This
  is similar to what has been proposed in~\citet{2014MNRAS.438.2466N} to
  estimate the period of rotating binary lens systems.}  For a binary lens
system, this does not even require the source to move parallel to its equator in
the lens plane, due to the fact the caustic is extended.  This technique is
already used for transit lightcurves, as tracking the change in position of a
starspot allows one to recover not only the orbital obliquity of the host star
with respect to the orbital plane, but also to get strong constraints on the
stellar rotation period (see, e.g.,~\citealt{2014MNRAS.443.2391M}).

A possible way to distinguish between a caustic crossing peak and a spot-induced
peak (\(c>1\)) or dip (\(c<1\)) is to observe the microlensing event in two
different bands.  Indeed, the contrast parameter \(c\) strongly depends on the
frequency of observation~\citep{2003ApJ...585L.147S} and the presence of a spot
characterized by a contrast parameter value much different than unity noticeably
affects the microlensing lightcurve, being this effect more evident at shorter
wavelengths.  Of course, we remark that observing such signatures requires high
precision and high cadence photometry and, possibly, observing the same target
with a network of dedicated telescopes (such as KMTNet, see,
e.g.,~\citealt{2012SPIE.8444E..47P}).

Finally, we remind that the optical depth for binary lens featuring star spots
is
\begin{equation}
  \tau_{\textup{s}} = f_{1} f_{2} f_{3} \tau
\end{equation}
where \(\tau\simeq 4.48\times10^{-6}\) is the microlensing optical depth towards
the Galactic bulge, \(f_{1} \simeq 0.184\) is the fraction of giant stars in the
Galactic bulge observed by OGLE-III~\citep{2015ApJS..216...12W}, \(f_{2} \simeq
0.054\) is the fraction of binary lenses in the OGLE-III data with
caustic-crossing features~\citep{2015ApJS..216...12W}, and, as assumed
by~\citet{2015MNRAS.452.2587S} (see also~\citealt{2005LRSP....2....8B}), \(f_{3}
\simeq 0.01\) is the fraction of giant stars with stellar spots induced by a
magnetic field stronger than \(100~\mathrm{G}\).  Here we are assuming that the
efficiency of detecting star spot features with residuals larger than
\(10^{-3}\) to be about unity, based on reachable photometric precision for
ground-based microlensing observations.  It turns out that \(\tau_{\textup{s}}
\simeq 5\times 10^{-10}\).  The number of events with star spots can be
estimated to be
\begin{equation}
  N_{\textup{s}} = \frac{\upi}{2} \frac{T_{\textup{obs}}
    N_{\textup{bg}}}{\langle t_{\textup{E}}\rangle} \tau_{\textup{s}}
\end{equation}
where \(N_{\textup{bg}} \simeq 400\cdot 10^{6}\) is the number of monitored
stars in the Galactic bulge by OGLE-IV~\citep{2015AcA....65....1U},
\(T_{\textup{obs}} = 365~\mathrm{d}\), \(\langle t_{\textup{E}}\rangle \simeq
24.6~\mathrm{d}\) \citep{2015ApJS..216...12W}.  Therefore, the number of events
with star spots features is about \(4.7\) per year.  We remark that the extra
peaks or dips are always within the typical crossing time of the caustic.

\section*{Acknowledgements}

We would like to thank the anonymous referee for his/her suggestions.  We
acknowledge the support by the INFN project TAsP.


\bibliographystyle{mnras}
\bibliography{bibliography}

\begin{thebibliography}{}
\makeatletter
\relax
\def\mn@urlcharsother{\let\do\@makeother \do\$\do\&\do\#\do\^\do\_\do\%\do\~}
\def\mn@doi{\begingroup\mn@urlcharsother \@ifnextchar [ {\mn@doi@}
  {\mn@doi@[]}}
\def\mn@doi@[#1]#2{\def\@tempa{#1}\ifx\@tempa\@empty \href
  {http://dx.doi.org/#2} {doi:#2}\else \href {http://dx.doi.org/#2} {#1}\fi
  \endgroup}
\def\mn@eprint#1#2{\mn@eprint@#1:#2::\@nil}
\def\mn@eprint@arXiv#1{\href {http://arxiv.org/abs/#1} {{\tt arXiv:#1}}}
\def\mn@eprint@dblp#1{\href {http://dblp.uni-trier.de/rec/bibtex/#1.xml}
  {dblp:#1}}
\def\mn@eprint@#1:#2:#3:#4\@nil{\def\@tempa {#1}\def\@tempb {#2}\def\@tempc
  {#3}\ifx \@tempc \@empty \let \@tempc \@tempb \let \@tempb \@tempa \fi \ifx
  \@tempb \@empty \def\@tempb {arXiv}\fi \@ifundefined
  {mn@eprint@\@tempb}{\@tempb:\@tempc}{\expandafter \expandafter \csname
  mn@eprint@\@tempb\endcsname \expandafter{\@tempc}}}

\bibitem[\protect\citeauthoryear{{Afonso}, {Alard}, {Albert}  et~al.}{{Afonso}
  et~al.}{2000}]{2000ApJ...532..340A}
{Afonso} C.,  {Alard} C.,  {Albert} J.~N.,   et~al., 2000, \mn@doi [ApJ]
  {10.1086/308561}, \href {http://adsabs.harvard.edu/abs/2000ApJ...532..340A}
  {532, 340}

\bibitem[\protect\citeauthoryear{{Berdyugina}}{{Berdyugina}}{2005}]{2005LRSP....2....8B}
{Berdyugina} S.~V.,  2005, \mn@doi [Living Reviews in Solar Physics]
  {10.12942/lrsp-2005-8}, \href
  {http://adsabs.harvard.edu/abs/2005LRSP....2....8B} {2, 8}

\bibitem[\protect\citeauthoryear{{Chang} \& {Han}}{{Chang} \&
  {Han}}{2002}]{2002MNRAS.335..195C}
{Chang} H.-Y.,  {Han} C.,  2002, \mn@doi [MNRAS]
  {10.1046/j.1365-8711.2002.05609.x}, \href
  {http://adsabs.harvard.edu/abs/2002MNRAS.335..195C} {335, 195}

\bibitem[\protect\citeauthoryear{{Gould}}{{Gould}}{2001}]{2001PASP..113..903G}
{Gould} A.,  2001, \mn@doi [PASP] {10.1086/322149}, \href
  {http://adsabs.harvard.edu/abs/2001PASP..113..903G} {113, 903}

\bibitem[\protect\citeauthoryear{{Gould}, {Yee}, {Bond}  et~al.}{{Gould}
  et~al.}{2013}]{2013ApJ...763..141G}
{Gould} A.,  {Yee} J.~C.,  {Bond} I.~A.,   et~al., 2013, \mn@doi [ApJ]
  {10.1088/0004-637X/763/2/141}, \href
  {http://adsabs.harvard.edu/abs/2013ApJ...763..141G} {763, 141}

\bibitem[\protect\citeauthoryear{Hahn}{Hahn}{2005}]{2005CoPhC.168...78H}
Hahn T.,  2005, \mn@doi [Computer Physics Communications]
  {10.1016/j.cpc.2005.01.010}, \href
  {http://adsabs.harvard.edu/abs/2005CoPhC.168...78H} {168, 78}

\bibitem[\protect\citeauthoryear{{Han}, {Park}, {Kim}  \& {Chang}}{{Han}
  et~al.}{2000}]{2000MNRAS.316..665H}
{Han} C.,  {Park} S.-H.,  {Kim} H.-I.,   {Chang} K.,  2000, \mn@doi [MNRAS]
  {10.1046/j.1365-8711.2000.03534.x}, \href
  {http://adsabs.harvard.edu/abs/2000MNRAS.316..665H} {316, 665}

\bibitem[\protect\citeauthoryear{Hendry, Bryce  \& Valls-Gabaud}{Hendry
  et~al.}{2002}]{2002MNRAS.335..539H}
Hendry M.~A.,  Bryce H.~M.,   Valls-Gabaud D.,  2002, \mn@doi [MNRAS]
  {10.1046/j.1365-8711.2002.05496.x}, \href
  {http://adsabs.harvard.edu/abs/2002MNRAS.335..539H} {335, 539}

\bibitem[\protect\citeauthoryear{{Heyrovsk{\'y}} \& {Sasselov}}{{Heyrovsk{\'y}}
  \& {Sasselov}}{2000}]{2000ApJ...529...69H}
{Heyrovsk{\'y}} D.,  {Sasselov} D.,  2000, \mn@doi [ApJ] {10.1086/308270},
  \href {http://adsabs.harvard.edu/abs/2000ApJ...529...69H} {529, 69}

\bibitem[\protect\citeauthoryear{{Kayser}, {Refsdal}  \& {Stabell}}{{Kayser}
  et~al.}{1986}]{1986A&A...166...36K}
{Kayser} R.,  {Refsdal} S.,   {Stabell} R.,  1986, A{\&}A, \href
  {http://adsabs.harvard.edu/abs/1986A%26A...166...36K} {166, 36}

\bibitem[\protect\citeauthoryear{{Mancini}, {Southworth}, {Ciceri}
  et~al.}{{Mancini} et~al.}{2014}]{2014MNRAS.443.2391M}
{Mancini} L.,  {Southworth} J.,  {Ciceri} S.,   et~al., 2014, \mn@doi [MNRAS]
  {10.1093/mnras/stu1286}, \href
  {http://adsabs.harvard.edu/abs/2014MNRAS.443.2391M} {443, 2391}

\bibitem[\protect\citeauthoryear{{Mao}}{{Mao}}{2012}]{2012RAA....12..947M}
{Mao} S.,  2012, \mn@doi [Research in Astronomy and Astrophysics]
  {10.1088/1674-4527/12/8/005}, \href
  {http://adsabs.harvard.edu/abs/2012RAA....12..947M} {12, 947}

\bibitem[\protect\citeauthoryear{{Nucita}, {Giordano}, {De Paolis}  \&
  {Ingrosso}}{{Nucita} et~al.}{2014}]{2014MNRAS.438.2466N}
{Nucita} A.~A.,  {Giordano} M.,  {De Paolis} F.,   {Ingrosso} G.,  2014,
  \mn@doi [\mnras] {10.1093/mnras/stt2363}, \href
  {http://adsabs.harvard.edu/abs/2014MNRAS.438.2466N} {438, 2466}

\bibitem[\protect\citeauthoryear{{Paczy\'{n}ski}}{{Paczy\'{n}ski}}{1986}]{1986ApJ...304....1P}
{Paczy\'{n}ski} B.,  1986, \mn@doi [ApJ] {10.1086/164140}, \href
  {http://adsabs.harvard.edu/abs/1986ApJ...304....1P} {304, 1}

\bibitem[\protect\citeauthoryear{{Park}, {Kim}, {Lee}, {Lee}  et~al.}{{Park}
  et~al.}{2012}]{2012SPIE.8444E..47P}
{Park} B.-G.,  {Kim} S.-L.,  {Lee} J.~W.,  {Lee} B.-C.,   et~al., 2012, in
  Society of Photo-Optical Instrumentation Engineers (SPIE) Conference Series.
  p.~47, \mn@doi{10.1117/12.925826}

\bibitem[\protect\citeauthoryear{{Queloz}, {Henry}, {Sivan}  et~al.}{{Queloz}
  et~al.}{2001}]{2001A&A...379..279Q}
{Queloz} D.,  {Henry} G.~W.,  {Sivan} J.~P.,   et~al., 2001, \mn@doi [A{\&}A]
  {10.1051/0004-6361:20011308}, \href
  {http://adsabs.harvard.edu/abs/2001A&A...379..279Q} {379, 279}

\bibitem[\protect\citeauthoryear{{Sajadian}}{{Sajadian}}{2015}]{2015MNRAS.452.2587S}
{Sajadian} S.,  2015, \mn@doi [\mnras] {10.1093/mnras/stv1349}, \href
  {http://adsabs.harvard.edu/abs/2015MNRAS.452.2587S} {452, 2587}

\bibitem[\protect\citeauthoryear{{Schneider} \& {Wei\ss}}{{Schneider} \&
  {Wei\ss}}{1986}]{1986A&A...164..237S}
{Schneider} P.,  {Wei\ss} A.,  1986, A{\&}A, \href
  {http://adsabs.harvard.edu/abs/1986A%26A...164..237S} {164, 237}

\bibitem[\protect\citeauthoryear{{Silva}}{{Silva}}{2003}]{2003ApJ...585L.147S}
{Silva} A.~V.~R.,  2003, \mn@doi [ApJL] {10.1086/374324}, \href
  {http://adsabs.harvard.edu/abs/2003ApJ...585L.147S} {585, L147}

\bibitem[\protect\citeauthoryear{{Skowron} \& {Gould}}{{Skowron} \&
  {Gould}}{2012}]{2012arXiv1203.1034S}
{Skowron} J.,  {Gould} A.,  2012, preprint (\mn@eprint {arXiv} {1203.1034})

\bibitem[\protect\citeauthoryear{{Udalski}, {Szyma{\'n}ski}  \&
  {Szyma{\'n}ski}}{{Udalski} et~al.}{2015}]{2015AcA....65....1U}
{Udalski} A.,  {Szyma{\'n}ski} M.~K.,   {Szyma{\'n}ski} G.,  2015, \actaa,
  \href {http://adsabs.harvard.edu/abs/2015AcA....65....1U} {65, 1}

\bibitem[\protect\citeauthoryear{{Witt}}{{Witt}}{1995}]{1995ApJ...449...42W}
{Witt} H.~J.,  1995, \mn@doi [ApJ] {10.1086/176029}, \href
  {http://adsabs.harvard.edu/abs/1995ApJ...449...42W} {449, 42}

\bibitem[\protect\citeauthoryear{Witt \& Mao}{Witt \&
  Mao}{1995}]{1995ApJ...447L.105W}
Witt H.~J.,  Mao S.,  1995, \mn@doi [ApJL] {10.1086/309566}, \href
  {http://adsabs.harvard.edu/abs/1995ApJ...447L.105W} {447, L105}

\bibitem[\protect\citeauthoryear{{Wolter}, {Schmitt}, {Huber}  et~al.}{{Wolter}
  et~al.}{2009}]{2009A&A...504..561W}
{Wolter} U.,  {Schmitt} J.~H.~M.~M.,  {Huber} K.~F.,   et~al., 2009, \mn@doi
  [A{\&}A] {10.1051/0004-6361/200912329}, \href
  {http://adsabs.harvard.edu/abs/2009A%26A...504..561W} {504, 561}

\bibitem[\protect\citeauthoryear{{Wyrzykowski} et~al.,}{{Wyrzykowski}
  et~al.}{2015}]{2015ApJS..216...12W}
{Wyrzykowski} {\L}.,  et~al., 2015, \mn@doi [\apjs]
  {10.1088/0067-0049/216/1/12}, \href
  {http://adsabs.harvard.edu/abs/2015ApJS..216...12W} {216, 12}

\bibitem[\protect\citeauthoryear{{Zechmeister} \& {K{\"u}rster}}{{Zechmeister}
  \& {K{\"u}rster}}{2009}]{2009A&A...496..577Z}
{Zechmeister} M.,  {K{\"u}rster} M.,  2009, \mn@doi [\aap]
  {10.1051/0004-6361:200811296}, \href
  {http://adsabs.harvard.edu/abs/2009A%26A...496..577Z} {496, 577}

\makeatother
\end{thebibliography}


\appendix
\section{Modelling star spots}
\label{sec:modelling-star-spots}

In this section we briefly describe the adopted modellization of the star spots.
We assume the spots to be perfectly circular on the surface of the star, but we
also take into account the effect of apparent distortion when they move towards
the star limb due to the star rotation.

In the sky plane, let \((r, \psi)\) be the polar coordinates of a generic point
on the star disc, with radius \(\rho\), then the three-dimensional Cartesian
coordinates of this point on the star surface are
\begin{align}
  x &= \rho\sqrt{1 - (r/\rho)^{2}}, \\
  y &= r\cos\psi, \\
  z &= r\sin\psi,
\end{align}
where the \(y-z\) plane coincides with the sky plane, and the \(x\)-axis is
directed towards the observer.  One can also calculate the colatitude \(\theta\)
and longitude \(\varphi\) of the point on the star surface with the usual
relations
\begin{align}
  \theta  &= \arccos(z/\rho), \\
  \varphi &= \arctan(y/x).
\end{align}
Let \((\theta_{i}, \varphi_{i})\) be the angular spherical coordinates of the
centre of the \(i\)th spot, with radius \(\rho_{i}\), on the star surface.  We
need a criterion to determine whether a point with coordinates \((\theta,
\varphi)\) on the star surface is inside a star spot.  For each spot, the
quantities
\begin{align}
  \theta'_{i}  &= \theta - (\theta_{i} - \upi/2), \\
  \varphi'_{i} &= \varphi - \varphi_{i}
\end{align}
are the angular spherical coordinates of the point \((\theta, \varphi)\) centred
on the centre of the spot.  If \(\cos\varphi'_{i} \ge 0\) then the spot and the
point are on the same side.  The projected two-dimensional Cartesian coordinates
of the point in this frame of reference are
\begin{align}
  y'_{i} &= \rho\sin\theta'_{i}\sin\varphi'_{i}, \\
  z'_{i} &= \rho\cos\theta'_{i}.
\end{align}
Thus, if the condition
\begin{equation}
  \sqrt{(y'_{i})^{2} + (z'_{i})^{2}} \le \rho_{i}
\end{equation}
is satisfied the point is inside the spot.  In this way, the effect of
distortion of the spots follows naturally.  We implemented this control in our
integration routine in order to set the contrast parameter.

\bsp 
\label{lastpage}
\end{document}